\documentclass[twocolumn,amsmath,amssymb,prl]{revtex4}
\usepackage{latexsym}
\usepackage{graphicx}
\usepackage{psfig}
\usepackage{color}
\usepackage[colorinlistoftodos]{todonotes}
\usepackage{verbatim}

\usepackage[letterpaper, margin=1in]{geometry}
\begin{document}

\newcommand{\note}{\textcolor{red}} \newcommand{\ceq}[1] {(\ref{#1})}

\title{Induced superconducting pairing in integer quantum Hall edge states} 

\author{Mehdi~Hatefipour$^{1}$}
\author{Joseph J. Cuozzo$^{2}$}
\author{Jesse Kanter$^{1}$}
\author{William M. Strickland$^{1}$}
\author{Christopher R. Allemang$^{3}$}
\author{Tzu-Ming Lu$^{3,4}$} \author{Enrico Rossi$^{2}$}
\author{Javad~Shabani$^{1}$} \email{jshabani@nyu.edu}

\affiliation{$^{1}$Center for Quantum Phenomena, Department of Physics, New York University,
NY 10003, USA \\
$^{2}$ Department of Physics, William \& Mary, Williamsburg, Virginia
23187, USA \\
$^{3}$ Sandia National Laboratories, Albuquerque, New Mexico 87185, USA \\
$^{4}$ Center for Integrated Nanotechnologies, Sandia National Laboratories, Albuquerque, New Mexico, 87123, USA }

\date{\today}

\begin{abstract} Indium Arsenide (InAs) near surface quantum wells (QWs)
are promising for the fabrication of semiconductor-superconductor heterostructures given that
they allow for a strong hybridization between the two-dimensional states in the quantum well
and the ones in the superconductor. In this work we present results for InAs QWs in the
quantum Hall regime placed in proximity of superconducting NbTiN. We observe a negative
downstream resistance with a corresponding reduction of Hall (upstream) resistance,
consistent with a very high Andreev conversion. We analyze the experimental data using the
Landauer-B\"{u}ttiker formalism, generalized to allow for Andreev reflection processes. We attribute the high efficiency of Andreev conversion in our devices to the large transparency of the InAs/NbTiN interface and the consequent strong hybridization of the QH edge modes with the states in the superconductor. \end{abstract}

\maketitle

Anyons with non-Abelian statistics  are of great fundamental interest \cite{MOOREREAD91} and
can be used to realize topologically protected, and therefore intrinsically fault-tolerant
qubits ~\cite{Kitaev2003, DasSarma05, Nayak2008}.
Non-Abelian anyons are expected to be realized in few fractional quantum Hall (QH) states
\cite{Read2000, Cheng2012, Alicea2016, Vaezi2013, Vaezi2014} such as the QH states with
filling factor $\nu=\frac{5}{2}$ \cite{Willett87, Miller07, Radu08}, and, possibly, $\nu =
\frac{12}{5}$ \cite{Zhu15}. However, so far, no  unambiguous experimental confirmation
exists of the presence of non-Abelian anyons in such QH states.

An alternative route to realize non-Abelian anyons relies on inducing superconducting
pairing between counter-propagating edge modes of QH states that, intrinsically, support
only Abelian anyons~\cite{Qi2010, Lindner2012,Clarke2013a, RogerMongPRX}. These theoretical
proposals build on an earlier proposal for creating Majorana zero modes, the anyons with the
simplest non-Abelian statistics, using 1D modes at the edge of a 2D topological insulator
(TI) in contact with a superconductor (SC)~\cite{FuKane2008}. In contrast to TIs, in
two-dimensional electron gases (2DEGs) in the QH regime, by varying filling factor $\nu$,
states can be realized with a variety of topological orders. This allows access to more
exotic edge states needed for engineering anyons with richer non-Abelian statistics. Key in
all these theoretical proposals is the ability to induce superconducting pairing, via the
proximity effect, between the QH edge modes.

The strength of the superconducting correlations that can be induced in a QH-SC
heterojunction can be evaluated by obtaining the amplitude of the Andreev reflection of QH
edge modes. The early search for Andreev reflection in QH-SC systems focused on InAs and
InGaAs semiconductor magneto-resistance oscillations at relatively low magnetic fields
\cite{Nitta_PRB_94} followed later by reports of induced superconductivity in QH states
\cite{Wan2015}. More recently there have been reports of observation of induced
superconductivity \cite{Rickhaus2012, Park2017}, cross Andreev conversion
\cite{gul2021induced,philip_kim_nature_IQH_2017}, edge state mediated supercurrent
\cite{Finkelstein_supercurrent}, and interference of chiral Andreev edge states
\cite{Finkelstein_nature_chiral_andreev,Hopppe_PRL_Andreev,Kurilovich2022,Manesco2021} in graphene. To make further progress, It is essential to reliably demonstrate the ability to induce robust superconducting correlations into the edge modes of a QH state.

In this work we show that in high quality InAs/NbTiN heterostructures, very strong  superconducting correlations can be induced in the edge modes of integer QH states realized  in the InAs-based quantum wells (QWs).
Such correlations appear to be robust, showing no oscillations as a function of doping, for gate voltages within the QH plateaus. We analyze the experimental data in conjunction with a microscopic model to extract the details of the processes determining the transport properties of the QH-SC interface.

\begin{figure*}[ht!] \centerline{\includegraphics[width=1.0\textwidth]{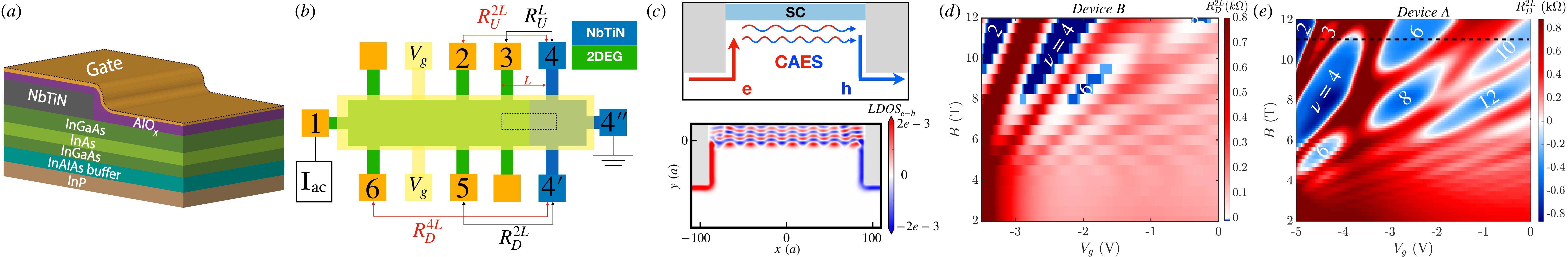}}
\caption{(a) Schematic of gated NbTiN/InAs hybrid device structure (only the portion which
is defined by the rectangle in Fig.~\ref{fig:fig1}(b)). (b) Device pin-out configuration.
Contacts  1,2,3,5, and 6 are normal; contacts 4, 4$^{\prime}$, 4$^{\prime\prime}$ are
superconducting. Contacts 1 and 4$^{\prime\prime}$ are used as the source and drain,
respectively. Contacts which are not labeled had electrical connection issue during the
experiment. (c) Andreev conversion via CAES interference along the QH-SC interface (top) and
a supporting tight binding calculation of the difference between the electron and hole LDOS
(LDOS$_{e-h}$) (bottom). (d) Measured $R_{D}^{2L}$ as a function of $V_g$ and $B$ in a dirty
interface device (device B). (e) Measured $R_{D}^{2L}$ as a function of $V_g$ and $B$ in the cleaned
interface device (device A). IQHSs are labeled from complementary $R_{xy}$ data. The dashed line shows the position of the cut shown in Fig.~\ref{fig:fig2} (a). } \label{fig:fig1} \end{figure*}

\begin{figure}[ht!] 
        \begin{center} \includegraphics[width=0.48\textwidth]{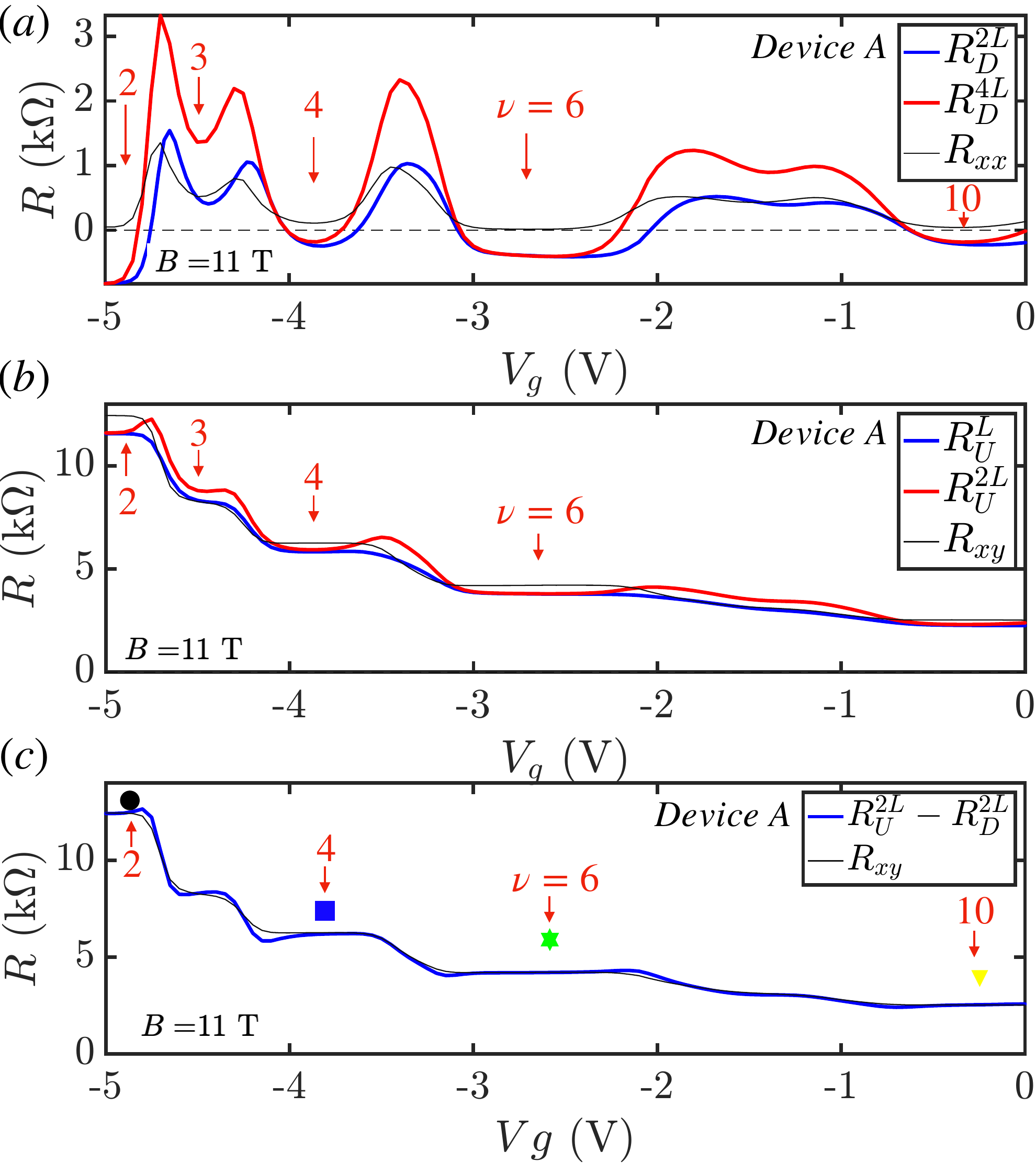}
        \caption{ 
$R_U-R_D$ and $R_{xy}$ shown as a function of $V_g$. All traces taken at $B=11$T (Device A). 
IQHSs are labeled and markers are shown for the states used in Fig.~\ref{fig:fig3}(c). 
        }
        \label{fig:fig2} 
    \end{center} 
\end{figure}

Figure~\ref{fig:fig1}(a) shows a cross sectional schematic of the fabricated device used in
this work. The QW is formed by a 4 nm layer of In$_{0.81}$Ga$_{0.19}$As layer, a 7 nm layer
of InAs, and a 10 nm top layer of In$_{0.81}$Ga$_{0.19}$As. The QW is grown on
In$_{x}$Al$_{1-x}$As buffer where the indium content is step-graded from $x =$ 0.52 to 0.81.
A delta-doped Si layer with electron doping $n\sim  1 \times 10^{12}$ cm$^{-2}$ is placed 6
nm below the QW.  This epitaxial structure has been used in previous studies on mesoscopic
superconductivity \cite{MortenNatureComm16, Bottcher2018, Mayer2019, Mayer2020}, in the
development of tunable qubits \cite{Casparis2019}, and in studies aimed at realizing and
detecting topological superconducting states~\cite{ SuominenPRL17, FornieriNature2019,
Dartiailh2021}. 

A Hall bar, Fig.~\ref{fig:fig1}(b), is fabricated by electron beam lithography. In order to
study the 2DEG/SC interface, a 90 nm thick layer of NbTiN was sputtered as the
superconducting contacts with a $150\mu m$-wide interface after performing wet etch surface
cleaning (Device A). We also fabricated a similar device with intentional no surface
cleaning step before NbTiN sputtering (Device B). A metallic top gate is created by
depositing a layer of Al oxide followed by an Al layer to control the QW electron
density~\cite{Shabani2016}.
The mobility of the QW is determined to be $\mu \sim 12,000$ $cm^2/V.s$ at $n \sim 8.51
\times 10^{11}$cm$^{-2}$ corresponding to an electron mean free path of $l_e \sim 180$~nm.
All data reported here were taken at $T \sim 30$~mK. We have provided more information on transport properties of the sample in the SI. We note that while we focus mainly on one device (Device A) in the main text, we have studied a few other similar devices which their data have been shown in SI.

When the sample is placed in a magnetic field, in the classical picture, electrons and holes
will alternate their skipping orbits across the interface of the superconductor and 2DEG
\cite{Chtchelkatchev2007}. In the full quantum-mechanical analysis the electron and hole
edge states hybridize due to the proximity of the SC and form a coherent chiral Andreev edge
state (CAES) extended along the QH-SC interface~\cite{Hopppe_PRL_Andreev, VanOstaay2011,
Khaymovich2010}. A schematic of CAES propagation along the QH-SC interface is shown in Fig.~\ref{fig:fig1} (c). 
In this picture, if more holes than electrons reach the
normal lead downstream from the superconducting electrode (lead 5), then a negative
potential difference $(V_5 - V_4')$ develops. In Fig.~\ref{fig:fig1} (c) we also show the
the local density of states of a CAES obtained with a tight binding (TB) calculation
performed using the python package Kwant \cite{Groth2014}. In the TB model the presence of
the magnetic field is taken into account via a Peierls phase, and the superconductivity of
the QW proximitized by NbTiN via a mean field 
an s-wave pairing term of strength $\tilde\Delta$. The details of the TB model can be found
in the SI.

Figures~\ref{fig:fig1}(d) and (e) show the results for the downstream resistance, $R_{D}^{2L}$, measured between the voltage contacts 5 and $4^{'}$,  as a function of gate voltage $V_g$ and magnetic field $B$. Hall resistance data measured between contacts 2 and 5 allow us to determine the filling factor of the different regions of Fig.~\ref{fig:fig1}~(d),~(e). Figure~\ref{fig:fig2}(a) shows the horizontal cut at $B=11$~T of Fig.~\ref{fig:fig1}(e), $R_{D}^{4L}$, and the corresponding longitudinal resistance $R_{xx}$. From the $R_{xx}$ measurements we see that we have well developed integer QH states (IQHS). From Figs.~\ref{fig:fig1}(e) and \ref{fig:fig2}(a) we clearly observe that $R_D$ is negative for IQHS, a fact that strongly suggests the presence of Andreev processes at the QH-SC
interface for these IQHS. We notice the importance of a clean InAs/NbTiN interface by comparing the magnitude of negative resistance in Figs.~\ref{fig:fig1} (d) and (e). The clean interface on device A has been achieved by etching the surface of defined NbTiN pattern area by buffered oxide etchant (BOE) for 2 seconds immediately followed by loading into sputtering tool's load lock in order to minimize the time for the native oxide growth at the interface. On the other hand, for device B and all the other devices mentioned in SI, this cleaning step has been skipped and NbTiN sputtered on the defined region after its exposure to air. For the rest of the paper, we focus only on device A results. The upstream resistance $R_U^{2L}$ (measured between
contacts 3 and 4) exhibits plateaus in correspondence to the $R_{xy}$ plateaus in magnetic
field but with resistance values lower than $R_{xy}$. Moreover,  $R_U^{2L}-R_D^{2L}$
recovers the quantized Hall value, $R_{xy}$, as shown in Fig.~\ref{fig:fig2}~(c). Note that
this difference does not necessarily match the $R_{xy}$ data outside the QH regime.

These results can be understood within the Landauer-B\"{u}ttiker (LB) theory, 
generalized to allow for the presence of a superconducting lead~\cite{Datta1996,Beconcini2018}.

We start with the six-terminal setup shown in Fig.~\ref{fig:fig1}(b) (see also the SI). We assume the terminal 1, 2, 3, 5, 6 to be ideal metallic leads, and contact 4 to be a superconducting lead. We first consider the limit in which no normal reflection or transmission processes take place at the superconducting lead. Let $I_i$, $V_i$, the currents and voltages, respectively, at the terminals $i=(1,2,3,4,5,6)$. Without loss of generality, we can set $V_4=0$. We can use the charge conservation equation $\sum_i I_i=0$ to express $I_4$ in terms of the currents at the other leads. With these considerations the LB equations reduce to the following system of linear equations:

    \begin{equation} 
    \begin{pmatrix} 
    I_1 \\ I_2 \\ I_3 \\ I_5 \\ I_6 
    \end{pmatrix}\!\! =\!\!
    \frac{\nu}{R_H}\! 
    \begin{pmatrix} 
    1  & 0  & 0 & 0 & -1 \\ 
    -1 & 1  & 0 & 0 & 0 \\ 
    0  &  -1 & 1 & 0 & 0 \\ 
    0  & 0  & 2A-1 & 1 & 0 \\ 
    0  & 0  & 0 & -1 & 1 
    \end{pmatrix}\!\!
    \begin{pmatrix} 
    V_1 \\ V_2 \\ V_3 \\ V_5 \\ V_6 
    \end{pmatrix} 
    \label{eq:conductanceM}
    \end{equation}

where $\nu$ is the number of edge states, $R_H$ is the Hall resistance, and $A$ is the
average probability, per edge mode, of Andreev reflection. 
Considering that no current flows into leads 2, 3, 5, 6, so that $I_2=I_3=I_5=I_6=0$, and $V_1=V_2=V_3$, $V_5=V_6$,
and setting $I\equiv I_1$, it is straightforward to solve Eq.~(\ref{eq:conductanceM}) to obtain
 
\begin{align} 
 &R_{U}^{2L} = \frac{V_2}{I} = \frac{R_H}{\nu} \frac{1}{2A} 
 \label{eq:upstream} \\ 
 &R_{D}^{2L} = \frac{V_5}{I} = \frac{R_H}{\nu} \left(\frac{1}{2A} - 1 \right) 
 \label{eq:downstream}
 \end{align}

\begin{figure}[ht!] 
 \begin{center} 
  \includegraphics[width=0.44\textwidth]{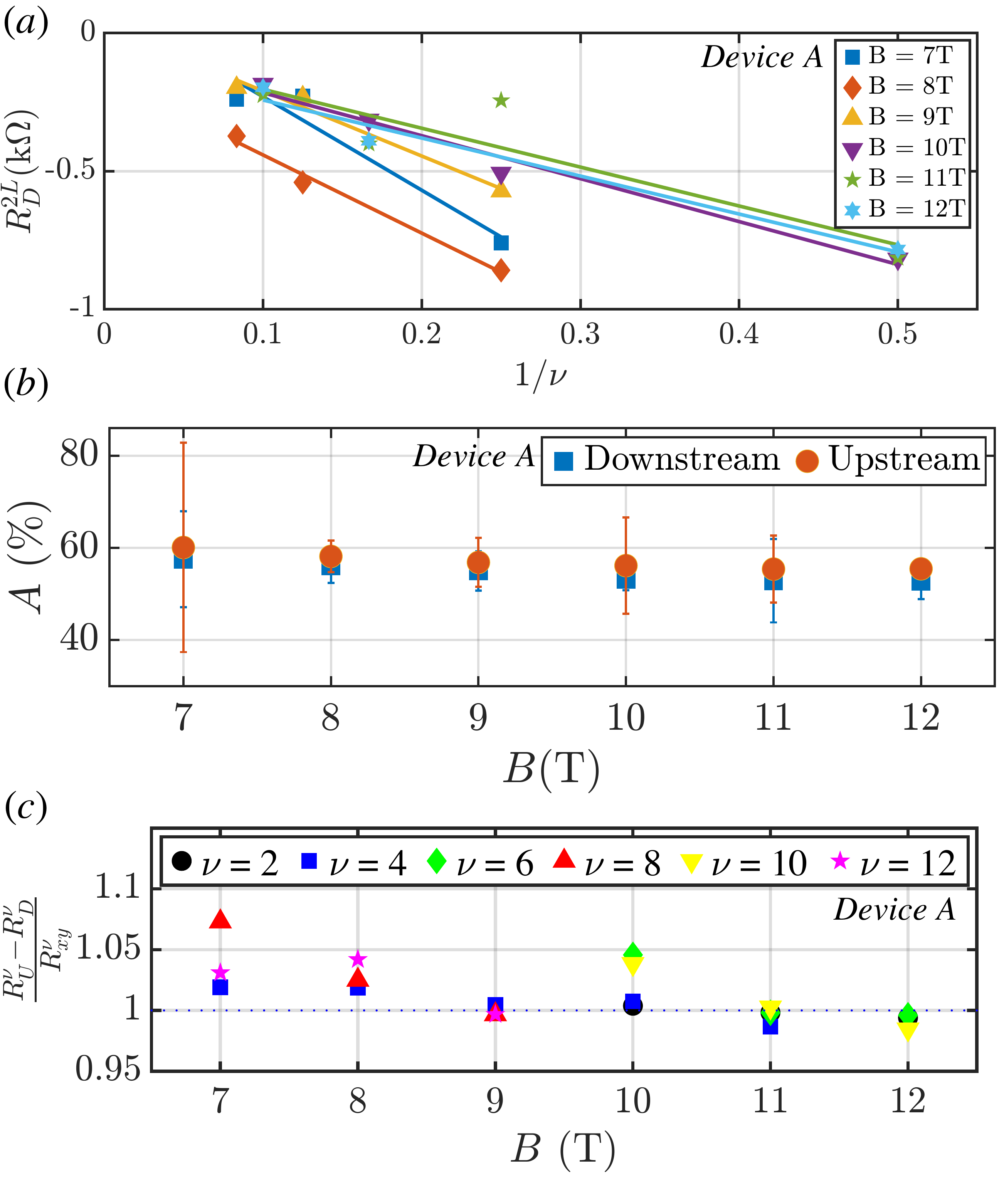}
  \caption{(a) $R_D^{2L}$  vs $1/{\nu}$ for different IQHSs and values of $B$, and linear fits
  corresponding to each magnetic field. (b) $A$ obtained from the slope of linear fits to
  $R_D$ and $R_U$ data vs $1/{\nu}$ with their corresponding error bars. (c)
  $(R_U^{2L}-R_D^{2L})/R^\nu_{xy}$  for different $\nu$s and values of $B$.} \label{fig:fig3}
 \end{center} 
\end{figure}

Figure~\ref{fig:fig3}~(a) shows the scaling of $R_D$ with respect to $1/\nu$ for different values of $B$.
From the slope of the fits to the experimental data shown in Fig.~\ref{fig:fig3}~(a) we obtain the value of $A_{\rm exp}=55\%$,
independent, to very good approximation, on the value of $V_g$ within the QH plateaus.
Figure~\ref{fig:fig3}~(c) shows the consistency of the measured values of $R_D^{2L}$ and
$R_U^{2L}$ with the LB predictions by plotting the ratio $(R_U^{2L}-R_D^{2L})/R_{xy}$ as
function of $B$ that, according to Eqs.~(\ref{eq:upstream}-\ref{eq:downstream}), is expected
to be equal to 1.

To understand qualitatively how such values of $A$ can arise, we consider
an effective 1D Bogoliubov de Gennes Hamiltonian, $H_{BdG}$, 
for the 1D chiral edge modes:
$H_{BdG} = \int dx~\psi^{\dagger}(x) \mathcal{H}(x) \psi(x)$ 
where $\psi(x) = (c_{x\uparrow},~c^{\dagger}_{x\downarrow})$ is the spinor formed by
the annihilation (creation) operator for a fermion at position $x$ and spin up (down)
and 
$\mathcal{H}(x) = v_{d}(-i\partial_{x})\tau_0 - v_d k_F \tau_z + \tilde{\Delta} \tau_x$. 
Here, and in the remainder, we set $\hbar=1$.
In the equation for $H(x)$, $v_d$ is the drift velocity of the edge modes, $\tau_i$ are the Pauli matrices 
Nambu space, $k_F$ is the edge modes Fermi wave vector 
(measured with respect to the QH-SC interface) 

and $\tilde\Delta$ is the superconducting pairing induced via the proximity effect by the superconducting lead.
Using the expression for $H$ we can obtain the transfer matrix $M$ relating $\psi(x)$ at the two ends of the 
length $L_{sc}$ of the QH-SC interface (see SI)~\cite{VanOstaay2011,Zhang2020},
and then the  expression for the electron-hole conversion
probability
\begin{equation}
 A = \frac{\sin^2(\delta \phi )}{[1 + (v_d k_F/\tilde{\Delta})^2]}.
 \label{eq:A}
\end{equation}
In Eq.~(\ref{eq:A}), $\delta\phi$ is the difference of the phases accumulated by the electron-like
and hole-like edge modes along the length of the QH-SC interface. 
Let $k_F^{(e)}$, $k_F^{(h)}$ be the Fermi wave vector of the electron-like and hole-like edge modes,
and $\delta k\equiv|k_F^{(e)}-k_F^{(h)}|=[\tilde\Delta^2+(v_d k_F)^2]^{1/2}/v_d$. 
We then write $\delta\phi=L_{sc}\delta k$.

Considering that $L_{sc}=150\mu{\rm m}$ is quite large any small change of $\delta k$, induced for example by changes in $V_g$,
should result in a significant change of $A$
and therefore of $R_U$ and $R_D$. However, in the experiment, within the QH plateaus, 
$R_U$ and $R_D$ do not show any oscillation as a function of $V_g$.
It is natural to conclude that this might be due to scattering processes leading
to a dephasing of the electron-like and hole-like modes along the QH-SC interface \cite{Kurilovich2022}. 
In this case the effective $A_{\rm eff}$ can be obtained by averaging over $L_{sc}$ on the right hand side of Eq.~(\ref{eq:A}) to obtain
$A_{\rm eff}=\langle A\rangle = (1/2){[1 + (v_d k_F/\tilde{\Delta})^2]}$.
Considering that $(v_d k_F/\tilde{\Delta})^2>0$ we see that in this case we cannot recover the value of $A$ extracted from the experimental results, $A=0.55>0.5$. 

To explain the large value of $A$, accompanied by the lack of oscillations as a function of $V_g$, we are led to two possibilities.
The first possibility is that $\delta_k$ does not change appreciably as a function of $V_g$. From Eq.~(\ref{eq:A}), considering that $0<\sin^2(\delta \phi)<1$, we can see
that to have $A=0.55$ we must have $v_d k_F/\tilde{\Delta}<0.9$. 
In the limit when $\delta\phi$ is such that $\sin^2(\delta\phi)\approx 0.55$, we must have
$v_d k_F/\tilde{\Delta}\ll 1$. In this limit we can 
write $\delta k\approx \tilde\Delta/v_d$. Considering that to good approximation, $\tilde\Delta$ and $v_d$
are independent of $V_g$, we recover the observed values of $R_U$ and $R_D$, with no oscillations, in the QH plateaus. Notice that the condition $v_d k_F/\tilde{\Delta}\ll 1$ is equivalent to the condition $\delta k \xi\approx  1$, where $\xi\equiv v_d/\Delta$ can be interpreted as the superconducting coherence length of the edge modes in proximity of the SC.
The other possibility is that dephasing processes are accompanied by a finite probability of single electron tunneling
into the superconductor and breaking of particle-hole (p-h) symmetry. 
This would allow to have a situation in which electron-like states are more likely than h-like 
states to tunnel into the superconductor and therefore contribute less to the downstream current
explaining a negative $R_D$ even when $A\leq 1/2$.
If we denote by $T$ the probability, per edge mode, of an electron-like state to tunnel in the SC,
in Eqs.~(\ref{eq:upstream}),~(\ref{eq:downstream}) we would replace $2A$ with $2A+T$. 
In this case from the measurements of $R_U$ and $R_D$ we recover $2A+T$.
Assuming $\langle A\rangle=1/2$ would imply $T=0.2$. The smallest value of 
$\langle A\rangle$, consistent with particle conservation, is $15\%$ to which it would correspond $T=0.8$,
a value that implies a very strong breaking of particle-hole symmetry at the QH-SC interface. It is difficult to distinguish between these two possibilities given that we cannot measure separately the quasiparticle and supercurrent contributions to the charge current flowing from the QH region into the superconducting lead.
\\
In conclusion, we have fabricated a quantum Hall-superconductor (QH-SC) epitaxial heterostructure based on InAs and NbTiN and characterized the transport properties of its QH edge modes propagating along a
superconducting interface. We have observed negative values for the downstream resistance
$R_D$ between a normal lead and the superconducting lead and a corresponding suppression of 
the upstream resistance $R_U$ such that in the QH plateaus the difference $R_U-R_D$
is equal to the Hall resistance $R_H$.
The negative values of $R_D$ are
an unambiguous sign that at the QH-SC interface there is a very large electron-hole
conversion probability, $A$. Using a Landauer-B\"{u}ttiker analysis we were able to
explain the relation between $R_D$ and $R_U$ and 
express both resistances in terms of a single effective probability for Andreev reflections at the QH-SC interface.
Our analysis led us to the conclusion that either the edge modes propagate along the QH-SC
interface with negligible dephasing resulting in an electron-hole conversion close to 55\%,
or, if dephasing processes dominate, that a strong breaking of particle-hole symmetry at the QH-SC interface must occur.

Even the lower bounds estimates for $A$ that we extract from our measurements are remarkable, larger
than any published results for QH-SC devices. This shows that in our InAs devices very
strong superconducting correlations 
can be induced into the QH edge modes, an essential
prerequisite to use QH-SC heterojunctions to realize non-Abelian anyons and topologically
protected qubits and quantum gates based on such unusual quantum states.

We thank Dr. Shashank Misra for his insightful comment and feedback. The NYU team
acknowledge partial support from DARPA grant no. DP18AP900007 and ONR  grant no.
N00014-21-1-2450. JJC and ER acknowledge support from ARO Grant No. W911NF-18-1-0290. JJC
also acknowledges support from the Graduate Research Fellowship awarded by the Virginia
Space Grant Consortium (VSGC). Sandia National Laboratories is a multimission laboratory
managed and operated by National Technology and Engineering Solutions of Sandia LLC, a
wholly owned subsidiary of Honeywell International Inc.~for the U.S.~DOE's National Nuclear
Security Administration under contract DE-NA0003525. This work was performed, in part, at
the Center for Integrated Nanotechnologies, a U.S.~DOE, Office of BESs, user facility.  This
paper describes objective technical results and analysis. Any subjective views or opinions
that might be expressed in the paper do not necessarily represent the views of the U.S. DOE
or the United States Government.

\bibliography{References_Shabani_Growth}

\pagebreak

\end{document}